\documentclass[aps,prd,10pt,twocolumn,superscriptaddress,showpacs,nobibnotes,longbibliography]{revtex4-1}

\usepackage{hyperref}
\usepackage{graphicx}
\usepackage{amsfonts,amsmath,amssymb,bm,bbm}
\usepackage{color,soul}
\usepackage{slashed}
\usepackage[toc,page]{appendix}
\usepackage{xcolor}

\hypersetup{
    pdfnewwindow=true,      
    colorlinks=true,       
    linkcolor=blue,          
    citecolor=blue,        
    filecolor=blue,      
    urlcolor=blue        
}

\begin{document}

\title{ Reverse Direct Detection: Cosmic Ray Scattering With Light Dark Matter}

\author{Christopher V. Cappiello}
\email{cappiello.7@osu.edu}
\thanks{\scriptsize \!\! \href{http://orcid.org/0000-0002-7466-9634}{orcid.org/0000-0002-7466-9634}}
\affiliation{Center for Cosmology and AstroParticle Physics (CCAPP), Ohio State University, Columbus, OH 43210}
\affiliation{Department of Physics, Ohio State University, Columbus, OH 43210}

\author{Kenny C. Y. Ng}
\email{chun-yu.ng@weizmann.ac.il}
\thanks{\scriptsize \!\! \href{http://orcid.org/0000-0001-8016-2170}{orcid.org/0000-0001-8016-2170}}
\affiliation{
Department of Particle Physics and Astrophysics, Weizmann Institute of Science, Rehovot, Israel}

\author{John F. Beacom}
\email{beacom.7@osu.edu}
\thanks{\scriptsize \!\!  \href{http://orcid.org/0000-0002-0005-2631}{orcid.org/0000-0002-0005-2631}}
\affiliation{Center for Cosmology and AstroParticle Physics (CCAPP), Ohio State University, Columbus, OH 43210}
\affiliation{Department of Physics, Ohio State University, Columbus, OH 43210}
\affiliation{Department of Astronomy, Ohio State University, Columbus, OH 43210} 

\date{3 February 2019}

\begin{abstract}
Sub-GeV dark matter candidates are of increasing interest, because long-favored candidates such as GeV-scale WIMPs have not been detected. For low-mass dark matter, model-independent constraints are weak or nonexistent. We show that for such candidates, because the number density is high, cosmic ray propagation can be affected by elastic scattering with dark matter. We call this type of search ``reverse direct detection," because dark matter is the target and Standard Model particles are the beam. Using a simple propagation model for galactic cosmic rays, we calculate how dark matter affects cosmic ray spectra at Earth, and set new limits on the dark matter-proton and dark matter-electron cross sections. For protons, our limit is competitive with cosmological constraints, but is independent. For electrons, our limit covers masses not yet probed, and improves on cosmological constraints by one to two orders of magnitude. We comment on how future work can significantly improve the sensitivity of cosmic-ray probes of dark matter interactions. 

\end{abstract}

\maketitle

\section{Introduction}
\label{sec:Introduction}

	Dark matter's (DM) particle properties, such as its mass and cross sections with Standard Model particles, are unknown, because the only conclusive evidence of DM is gravitational \cite{Bertrev04, Bau12, Pet12,BucRev,Ber16,Bau18}. As commonly considered DM candidates such as GeV-scale Weakly Interacting Massive Particles (WIMPs) have not been found after many years of searching \cite{dd8, dd9, dd6, Kahrev, ATLAS17, CMS18, Lea18}, interest in more general DM candidates has grown \cite{Mac07, Alb10, Ess12, Ess15, Ali15, Ang17, An18, Emk18,ADMX18}.

One example, which we focus on here, is DM with mass $m_{\chi}$ in the keV--GeV range. Still-lower masses are disfavored (except for bosonic DM, such as axions) due to their effects on structure formation \cite{Tre79, Aba06,Vie06, Boy08, Pet12, Hor14}. For masses $\lesssim$ 1 GeV, present constraints are much weaker than direct-detection limits on GeV-scale DM. For such low masses, the energy transfer measured by direct detection experiments is small compared to typical detector thresholds; in indirect detection, many Standard Model states are below threshold; and in collider searches, there is a ceiling in cross section that gets low for small $m_{\chi}$ (see below). For masses $\lesssim$ 1 GeV, the tightest constraints come from cosmological and astrophysical tests, which gain sensitivity with increased number density.

\begin{figure}[t!]
\centering
\includegraphics[width=\columnwidth]{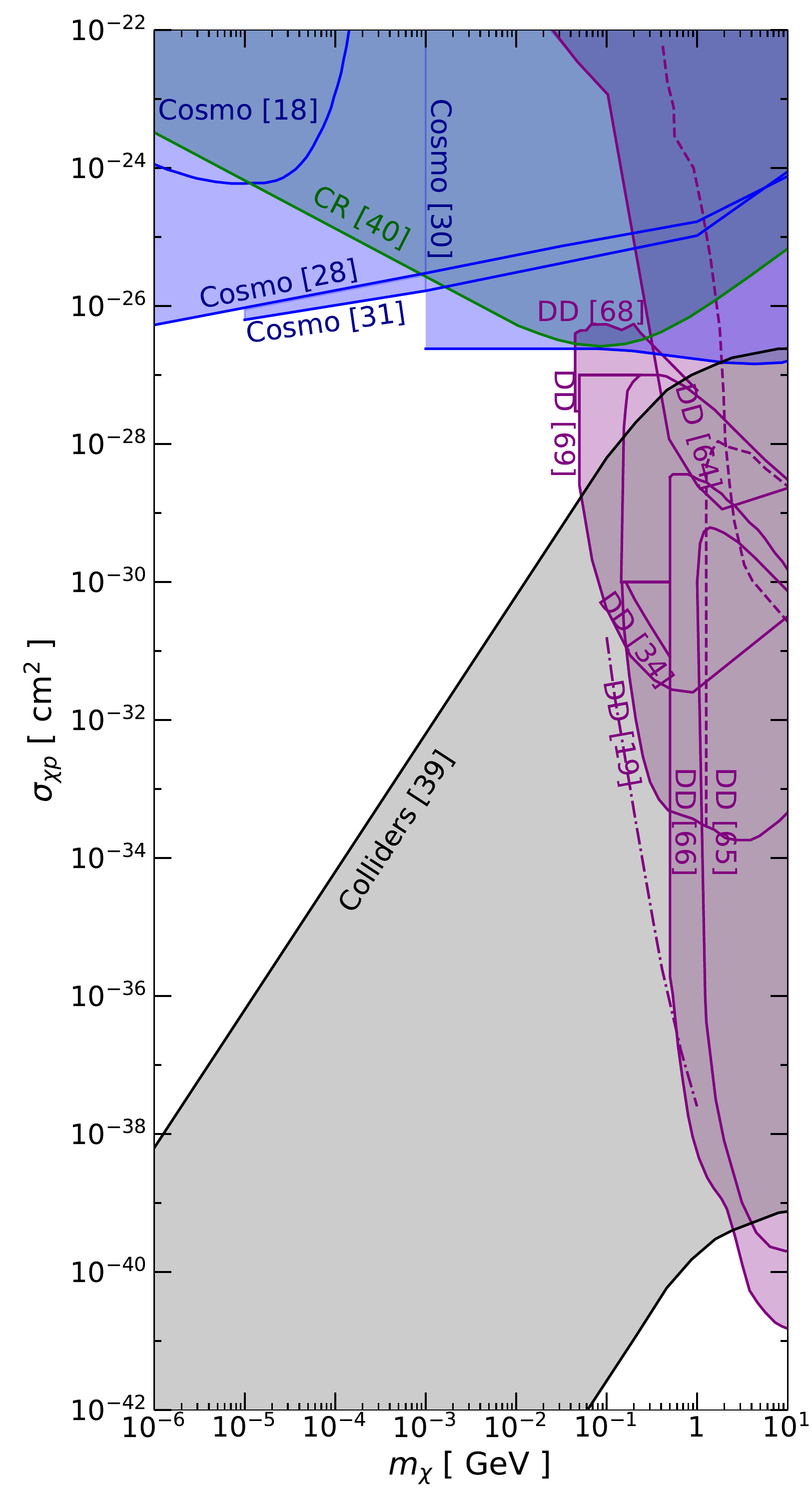}
\caption{Exclusion regions for DM-proton scattering from cosmology \cite{Glu17, Ali15, Xu18, Sla18}, colliders \cite{Dac15}, CRs \cite{Cyb02}, and direct detection with XQC \cite{Eri07}, DAMIC (\cite{DAMIC16}; ceiling from \cite{Mah18}), CRESST-II (\cite{dd4}; ceiling from \cite{Emk18a}), the CRESST surface run \cite{Ang17}, XENON100 \cite{Kou17}, EDELWEISS \cite{EDELWEISS}, and a near-surface detector at the University of Chicago \cite{Col18, Awe18}.  The dashed curves are reanalyses of XQC and the CRESST surface run from Ref.~\cite{Mah18}. The dash-dotted curve is a reanalysis of XENON1T from Ref.~\cite{Dol18}.} 
\label{fig:prevlims}
\vspace{-0.25 cm}
\end{figure}

For DM-proton interactions, cosmological limits on the scattering cross section for low-mass DM require $\sigma \lesssim 10^{-27}$ cm$^2$ \cite{Glu17,Bod18,Xu18, Sla18}. And even if direct-detection experiments improve their recoil sensitivity to probe lower masses (see Ref.~\cite{bat17} and references therein), underground experiments are likely only sensitive to $\sigma \lesssim 10^{-30}$ cm$^2$ due to their overburden \cite{Col93, Kou17, Mah18}.  Collider missing-momentum searches set limits on the DM-proton coupling $G$ to a heavy mediator, which can be translated into strong constraints on the scattering cross section that scale as $\sigma_{\chi p} \propto G^2 \mu_{\chi p}^2 \sim G^2 m_{\chi}^2$ \cite{Abd15, Bov16, Bov18}.  However, there is a cross section ceiling that also scales as $m_{\chi}^2$, above which the DM would interact in the detector, failing to register as missing energy \cite{Dac15}. Figure~\ref{fig:prevlims} summarizes the situation, showing a large, unconstrained region of low-mass DM parameter space between cosmological probes and collider limits. Additional constraints can be derived from the heating of gas clouds \cite{Chi90, Bho18}, which is relevant at higher masses. With some model dependence, constraints can also be obtained by considering Casimir-Polder type forces between nucleons \cite{Fic18, Bra18}, cooling of stars and supernovae \cite{Raf99, Fay06, Pos08, Raf08, Guh18}, astrophysical observations \cite{Brd18, Nam18}, and the early universe \cite{Ste13, Boe13, Nol15}.

For DM-electron interactions, the situation is similar.  Cosmological probes have only constrained the cross section to be $\sigma \lesssim 10^{-27}$ cm$^2$, but only for $m_{\chi} \lesssim 1$ MeV \cite{Ali15}. There should be a sensitivity ceiling for collider searches, though its value has not been determined.  

New ideas are needed for model-independent probes of sub-GeV DM. While direct detection ($\chi + SM \rightarrow \chi + SM$), indirect detection ($\chi + \chi \rightarrow SM + SM$), and collider searches ($SM + SM \rightarrow \chi + \chi$) are well-known types of DM search, what we call ``reverse direct detection" ($SM + \chi \rightarrow SM + \chi$), in which a Standard Model beam scatters with a near-stationary DM target, is less often considered. One example is CRs scattering with DM as they propagate. Past studies have considered inelastic interactions of DM with CR protons \cite{Cyb02, Cha12, Hoo18, Bey18} or particles in AGN jets \cite{Blo98, Gor10, Cha12, Pro13} that produce gamma rays or neutrinos. Scattering from laboratory beams has also been considered \cite{Kah08, Neu18}, as has scattering of cosmic neutrinos with DM \cite{Arg17}. Cosmological studies could also be considered an example of reverse direct detection, although both DM and SM particles are the beam and target, as they are in thermal motion.

	We propose a new method of reverse direct detection that probes low-mass {\bf DM-proton} and {\bf DM-electron} interactions by considering Milky-Way cosmic rays (CRs) \textit{elastically scattering} with DM. If CR protons and electrons scatter with DM particles as they propagate in the galaxy, they will lose energy in the collisions, and these losses will alter the observed CR spectra. This effect is most important at low DM mass. Using a simplified galactic CR propagation model, we calculate how scattering with DM affects the CR spectra, and set limits for protons that are competitive with current constraints and limits for electrons that reach a previously unconstrained mass range and which improve upon existing constraints by one to two orders of magnitude. Significant improvements in sensitivity are possible, as discussed below.

This paper is organized as follows. In Sec.~\ref{sensitivity}, we describe the basics of our proposal and estimate its sensitivity. In Sec.~\ref{propagation}, we calculate the energy loss rate of CRs through elastic scattering with DM, and present our model of CR propagation. In Sec.~\ref{protons}, we calculate results for protons. In Sec.~\ref{electrons}, we do the same for electrons. In Sec.~\ref{discussion}, we discuss additional considerations. In Sec.~\ref{conclusions}, we present our conclusions.

\section{Sensitivity Estimates}
\label{sensitivity}

Starting from the basic properties of CRs in the galaxy, we estimate the sensitivity reach of DM-CR scattering to the DM-proton and DM-electron scattering cross sections, showing that interesting regions of parameter space can be probed. For these estimates, we focus on CR protons and electrons with energies of roughly 10 GeV. Below, we use a range of energies for which the sources are certainly galactic \cite{Bla13}. Importantly, due to galactic magnetic fields, CRs do not travel in straight lines, and their motion is typically described as diffusion. For this reason, CRs are confined in the galaxy for much longer than the light-crossing time, greatly increasing their expected number of collisions with DM. For simplicity, we neglect Standard Model energy losses in this section.

To estimate the cross section above which DM has a noticeable impact on the CR spectrum, we need to know how long CRs propagate in the galaxy. The escape time can be determined from measurements of radioactive secondaries, and is roughly 15 Myr $\times \left(\frac{R}{10\,\textrm{GV}}\right)^{-\delta}$, where $\delta \simeq 0.6$ and $R$ is the rigidity ($R = p/|q|$) \cite{CRIS}. For highly relativistic particles, $p \simeq E_{\rm kinetic}$, so the escape time is often approximated as a power law in kinetic energy, but we use rigidity to be precise. Throughout this paper, we use $E$ to denote kinetic energy. We write the distance traveled by a CR as $L = c\, T_{\rm esc}$, where $c$ is the speed of light, which we display explicitly in some cases to make units more clear. Assuming that the average DM density experienced by the CR is roughly the local density of $\rho_{\rm DM} = m_{\chi}n_{\rm DM} =$ 0.3 GeV cm$^{-3}$, we estimate the cross section that corresponds to a single CR interaction using $n \sigma L \sim 1$, leading to $\sigma \sim m_{\chi} / (\rho_{\rm DM} L)\,$, or 
\begin{equation}\label{eq:1int}
\sigma_{\rm 1-int} \sim 2\times 10^{-31}\,{\rm cm^{2}} \left( \frac{m_{\chi}}{\rm keV} \right) \left( \frac{E}{\rm 10\,GeV} \right)^{\delta}\, .
\end{equation}

Equation~(\ref{eq:1int}) sets a scale, but is overly simplistic, as many collisions may be required to appreciably affect CR observables.  A more realistic estimate can be obtained by considering CR energy loss.  The energy loss in one DM-CR elastic collision is 
\begin{equation}\label{eq:eloss}
|\Delta E|  =  \Delta E_{\rm max} \frac{(1-\cos\theta)}{2}\, ,
\end{equation}
where $\theta$ is the center-of-momentum (CM) scattering angle and $\Delta E_{\rm max}$ is the maximum kinematically allowed energy transfer (see Sec.~\ref{propagation}).

So far, this is general to protons and electrons, which differ only in scattering kinematics. For protons with $E \ll m_{p}^{2}/2 m_{\chi}$, the angular-averaged fractional energy loss per collision, $\langle |\Delta E| \rangle/E \simeq m_{\chi} E/m_p^2$ is much smaller than unity. Thus, many DM-CR interactions are needed to appreciably affect the CR's energy. Taking this into account, we estimate the cross section at which a CR would lose roughly all of its energy, 
\begin{align}
\sigma^{\chi p}_{\rm loss} &=  \sigma_{\rm 1-int}\frac{E}{\langle |\Delta E| \rangle}\\ 
&\sim \sigma_{\rm 1-int} \left(\frac{m_p^2}{m_{\chi}E}\right) \\ \label{eq:elossxsec}
&\sim 2 \times 10^{-26} \left(\frac{E}{10 \, \rm GeV} \right)^{\delta-1} \, {\rm cm}^2\, ,
\end{align}
At low $m_{\chi}$, the energy loss per collision is proportional to $m_{\chi}$, so that the number of collisions needed to cause a DM particle to lose all its energy scales as $1/m_{\chi}$. This cancels the factor of $m_{\chi}$ in $\sigma_{\rm 1-int}$ that comes from the number density, so that for the lowest DM masses that we consider the energy loss rate is independent of DM mass.

We estimate our sensitivity for CR electrons using the same framework. For electron energies around 10 GeV, escape is the dominant loss process (see Sec.~\ref{electrons}), so we neglect Standard Model energy losses for this estimate. Because CR electrons are extremely relativistic, unlike CR protons, $\langle |\Delta E| \rangle/E \sim \frac{1}{2}$, so that
\begin{align}
\sigma^{\chi e}_{\rm loss} &=  \sigma_{\rm 1-int}\frac{E}{\langle |\Delta E| \rangle} \\ &\sim 2\, \sigma_{\rm 1-int} \\ \label{electronxsec} &\sim 4 \times 10^{-31} \left(\frac{m_{\chi}}{\rm keV} \right) \left(\frac{E}{10 \,\rm GeV} \right)^{\delta} \, {\rm cm}^2\, .
\end{align}
much tighter than current cosmological constraints \cite{Ali15}. 

Based on these estimates, our method is promising, especially at low DM mass. A more careful investigation is thus warranted to explore its reach. In the rest of this paper, we study how DM-CR scattering can affect the CR spectrum using a simple CR propagation model.   

\section{CR Propagation without and with DM Interactions}
\label{propagation}

We briefly overview CR propagation in the absence of DM. We then give the general energy loss rate of CRs due to scattering with DM, which will allow us to model CR propagation in the presence of DM scattering. 

\subsection{CR Propagation}
	
	The propagation of CRs in the galaxy can be described by a diffusion equation \cite{Gin64, Gai90, Str07}:
\begin{equation}\label{diffusion}
\begin{aligned}
&\frac{dN(E)}{dt} - \nabla\cdot[D(E)\nabla N(E) + {\bf V}N(E)]+ \frac{d}{dE}\left[\frac{dE}{dt}N(E)\right] \\&=
Q(E) - \frac{c\rho \sigma}{\lambda} + \sum_k \int_E^{\infty} dE' \frac{d\sigma_{k}(E',E)}{dE}n_k(E').
\end{aligned}
\end{equation}
Here the first term is the time derivative of the CR spectrum $N(E)$, the second term represents diffusion and advection with coefficient $D(E)$ and advective velocity ${\bf V}$, the third term represents energy loss and gain, $Q(E)$ is the source production spectrum, and the last two terms represent loss due to collisions with the interstellar medium and secondary CR production from spallation and decay of other CR species, $k$. We neglect diffusive reacceleration.

We require steady-state solutions, so the time derivative term is set to zero. And we consider only proton and electron CRs, for which the collision loss and spallation production terms are negligible. In the energy range we consider, the baryon grammage is small and we neglect CR collisions with the interstellar medium. We thus obtain
\begin{equation}\label{diffusion2}
\begin{aligned}
- \nabla\cdot[D(E)\nabla N(E) + {\bf V}N(E)] + \frac{d}{dE}\left[\frac{dE}{dt}N(E)\right] = Q(E).
\end{aligned}
\end{equation}

To further simplify this, we replace the diffusion-convection term with a term representing escape from the galaxy, $\frac{N(E)}{T_{\rm esc}(E)}$, where $T_{\rm esc}$ is the escape time. This simplification is known as the Leaky Box Model~\cite{Fer54, Gai90, Str07}. So finally, we reduce Eq.~(\ref{diffusion}) to the Leaky Box Equation:
\begin{equation}\label{LeakyBox}
\frac{N(E)}{T_{\rm esc}} + \frac{d}{dE}\left[\frac{dE}{dt}N(E)\right] = Q(E).
\end{equation}
The escape time's dependence on rigidity is parametrized as $T_{\rm esc} = 15\, \textrm{Myr} \times (\textrm{R}/10\, \textrm{GV})^{-\delta}$ for rigidities in our energy range, with $\delta = 0.58$, according to an analysis by the CRIS collaboration \cite{CRIS}. The CRIS collaboration fits a break in the rigidity dependence of the escape time at 1.4 GV; other analyses have placed this break between one and several GV \cite{Nat12, Jon01, Gai90}. We restrict our analysis to energies above 10 GeV ($R \simeq 10\,$ GV), so that we can treat the escape time as a power law in rigidity, and so we can also safely neglect solar modulation.

We use Eq.~(\ref{LeakyBox}) to model proton and electron propagation separately. The source spectra for protons and electrons are different, and the energy loss terms differ as well: Standard Model energy losses are small for protons of all energies, while for electrons, synchrotron and inverse-Compton losses become important above $\sim$100\,GeV. For both protons and electrons, we consider energy loss due to collisions with the interstellar medium to be negligible. To be conservative, we allow $\delta$ to vary independently for the two cases.

\subsection{CR Energy Loss from DM Collisions}

In the presence of DM-CR scattering, an additional energy-loss process affects CR propagation. Since we consider elastic scattering, where particle number is conserved, we can incorporate this effect into the energy-loss term $dE/dt$. The same effect could be achieved through the more general particle loss and production terms, but for our case this is unnecessary. In the continuous limit, the DM induced energy loss rate is
\begin{equation}\label{DMloss}
\frac{dE}{d t} = c\frac{\rho_{\rm DM}}{m_{\chi}} \int_0^{\Delta E_{\rm max}}dK\; K \frac{d\sigma}{dK}\, ,
\end{equation}
where $\rho_{\rm DM}$ is the DM density, $c$ is the speed of light, $d\sigma/dK$ is the differential cross section as a function of the final DM kinetic energy, $K$, and $\Delta E_{\rm max}$ is the maximum energy transfer for two-body elastic scattering~\cite{Sch02}. $\Delta E_{\rm max}$ is obtained from Ref.~\cite{Jac98} as
\begin{equation}\label{deltaE}
\Delta E_{\rm max} = \frac{ 4 m_{\rm CR}(1+\frac{E}{2m_{\rm CR}}) \frac{E}{m_{\chi}} }{(1+\frac{m_{\rm CR}}{m_{\chi}})^2+ \frac{2E}{m_{\chi}}}\,.
\end{equation}
There are two kinematic regimes, depending on the relative importance of the two terms in the denominator. Which is most relevant depends on the CR particle and energy and the DM mass, as we detail in the next two sections.

For simplicity, we assume the DM scattering cross section, $\sigma_{\chi p}$ or $\sigma_{\chi e}$, is energy independent in the energy range of interest, and that $d\sigma/dK$ is a flat distribution, representing isotropic scattering in the CM frame.  We comment on these assumptions in Sec.~\ref{discussion}. We then have $d\sigma/dK = \sigma/\Delta E_{\rm max}$.  
For either proton or electron CRs, the energy loss rate is then
\begin{eqnarray}
\label{Eqa}
\frac{dE}{dt} &=& c \frac{\rho_{\rm DM}}{m_{\chi}} \sigma \frac{ \Delta E_{\rm max} }{2}\, .
\end{eqnarray}
We take the average galactic DM density to be the local value of $\rho_{\rm DM} \simeq \,\,$0.3 GeV/cm$^3$ \cite{Web10}. This is conservative: although CRs propagate in a halo that extends outside the galactic disk, their sources are more concentrated in the inner galaxy, where the average DM density is higher.


\section{Constraining The DM-Proton Cross Section, $\sigma_{\chi p}$}
\label{protons}

As shown in Figure~\ref{fig:prevlims},  a large window exists at low DM mass and moderately large cross sections between the regions probed by cosmology and colliders. Direct-detection experiments can only probe larger masses, though some exclusion regions have been extended by considering DM acceleration through solar reflection (\cite{Emk18}; Ref. \cite{An18} considered the same effect for electron scattering). We show how DM-CR proton scattering can constrain this region.

\subsection{Proton Data and Model Without DM}

We begin by modeling proton propagation in the absence of DM interactions in the energy range from 10 GeV to 200 TeV. We model the proton source spectrum as a broken power law in rigidity, with spectrum indices $\gamma_1$, $\gamma_2$ and $\gamma_3$, with breaks corresponding to the observed breaks at around 500 GeV and 20 TeV. We model the escape time as a power law in rigidity, $T_{\rm esc} = 15\, \textrm{Myr}\times (R/10\, \textrm{GV})^{-\delta}$. Our model has seven fit parameters: $\delta$, $\gamma_1$, $\gamma_2$, $\gamma_3$, the break energies $E_{B1}$ and $E_{B2}$, and the normalization $Q$ of the source spectrum (denoted $Q_0$ in the no-DM case). Without energy-loss terms, the solution for the spectrum is trivial: 
\begin{equation}\label{noDMsol}
N(E) = Q(E)\,T_{\rm esc}(E).
\end{equation}
There is degeneracy between \{$\gamma_1$, $\gamma_2$, $\gamma_3$\} and $\delta$: for $Q(E) \propto E^{-2.2}$ and $T_{\rm esc}(E) \propto E^{-0.5}$, the observed spectrum is $N(E) \propto E^{-2.7}$, as observed for CR protons. But a harder source spectrum and steeper escape term could produce the same spectrum. This degeneracy is broken with the inclusion of DM, as the solution for $N(E)$ is no longer trivial, so we include all seven parameters here for completeness.

We assume that CRs in our energy range of interest are accelerated in galactic supernova remnants. This paradigm dates back to the 1930s \cite{Baa34}, and today remains the most likely explanation of CR acceleration, being supported by several lines of evidence \cite{Bla11,Bla13,Mor17,Byk18}. Supernovae are the only galactic sources with enough energy to explain the observed CR flux; it is estimated that $\sim$10\% of supernovae's kinetic energy must go into CRs to account for the CR spectrum seen at Earth \cite{Gin64, Str10, Bla11, Der13, Bla13}. Diffusive shock acceleration in supernova remnants is also expected to produce a CR source spectrum somewhat steeper than $E^{-2}$, which when combined with an escape time with an exponent of $\delta = 0.5 - 0.6$ \cite{CRIS, Obe12} can produce the observed $E^{-2.7}$ spectrum. And recently, observations of hadronic gamma rays from multiple supernova remnants have provided direct evidence for acceleration of relativistic protons in supernova remnants, with inferred CR acceleration efficiency that is roughly consistent with the 10\% mentioned above \cite{Giu11, Fermi13}.

The assumption of acceleration in supernova remnants provides two restrictions on our fit parameters. First, for diffusive shock acceleration in supernova remnants, the source spectrum cannot be harder than $E^{-2}$ (see, e.g., Ref.~\cite{Bla13}). Second, the energy injected into CRs cannot be arbitrarily high. As described above, collisions with DM will cause CRs to lose energy; such energy loss could be compensated by increasing the total energy in the source spectrum. However, as the average supernova kinetic energy and supernova rate in the galaxy are well known, the only freedom we have is to increase the CR acceleration efficiency. This efficiency is uncertain, but if it must be $\sim$10\% to account for the observed CR flux, then letting this efficiency change clearly cannot increase the total energy by a factor of more than $\sim$10.

We use the CR proton-only data measured by AMS \cite{AMSproton} and CREAM-I + CREAM-III \cite{CREAM} over the energy range 10\,GeV to 200\,TeV. (We believe that the error bar reported in the 6.31--10 TeV bin of the CREAM data has a typo and should be larger by a factor of 10, and have made this correction.) We include both statistical and systematic uncertainties, and treat the systematic uncertainties in different energy bins as uncorrelated, for simplicity and to be conservative. These data cover the largest range of energies with the smallest uncertainties; for other data, see Refs.~\cite{ATIC,PAMELA}. Below the specified energy range, solar modulation becomes important. And above this range, the CR spectrum steepens around 1 PeV, and the uncertainties in both observation and theory increase. We choose not to use the all-particle CR spectrum, as additional considerations are needed for heavier species of nuclei; see Sec.~\ref{discussion} for discussion.

Figure~\ref{fig:badfit} (bottom panel) shows the CR data and our best-fit model without DM. 
The values of the fit parameters are $\{\delta, \gamma_1, \gamma_2, \gamma_3, E_{B1}, E_{B2}
\} = \{0.5,2.4, 2.1, 2.4, 540\, \rm{GeV}, 23000\, \rm{GeV}\}$, all reasonable values. The normalization $Q$ of the source spectrum is consistent with $\sim10\%$ of supernova kinetic energy going into CRs, as discussed above. We will refer to the best fit source spectrum without DM scattering as $Q_0(E)$. 
Our model in the absence of DM is an excellent fit to the data, with a $\chi^2$ per degree of freedom of 0.25. The small $\chi^2$ value is likely because of our conservative choice to treat the systematic uncertaintiess as uncorrelated.  Overall, this shows that the data can be well described by a broken power law.

\subsection{Proton Spectrum with DM-Proton Scattering}	

The effect of DM-proton scattering comes in the form of additional energy loss during CR propagation. We quantify the significance of the DM energy loss term at a given energy by defining $T_{\rm loss} = E/|dE/dt|$ as the characteristic timescale for CRs to lose energy due to scattering with DM. For protons, the DM energy loss rate given by Eq.~(\ref{Eqa}) is
\begin{align}
\frac{dE}{dt} &= c\, \frac{\rho_{\rm DM}}{m_{\chi}}\, \sigma \,  \frac{m_{\chi}\left(2m_p E + E^2\right)}{(m_{\chi} + m_p)^2 + 2m_{\chi}E} \,.
\end{align}
We note that there are two different kinematic regimes here.  For small DM mass and low CR energy, $2 m_{\chi}E \ll (m_{p}+ m_{\chi} )^2$, $dE/dt \propto E^{2}$ and is independent of the DM mass. The latter fact is due to the energy loss per collision, Eq.~(\ref{eq:eloss}), being proportional to $m_{\chi}$, which cancels the $1/m_{\chi}$ factor due to DM number density.  In the opposite limit, where $2 m_{\chi}E \gg (m_{p}+ m_{\chi} )^2$, $dE/dt \propto E/m_{\chi}$, which causes our method to lose power at high masses.

	We determine the effects of DM-CR scattering on the CR spectrum by solving the Leaky Box Equation as a differential equation in energy for the spectrum $N(E)$. If the cross section is very small, the spectrum approaches the no-DM solution, Eq.~(\ref{noDMsol}). For arbitrary cross section, the solution of the Leaky Box Equation is \cite{Arf05}
\begin{equation}\label{LBSol}
\begin{aligned}
N(E) = &\int_E^{\infty}dE'\,\frac{Q(E')}{dE'/dt} \\
&\times \exp\left(-\int_E^{E'}\frac{dE''}{(dE''/dt) T_{\rm esc}(E'')}\right).
\end{aligned}
\end{equation}
This same equation has been used in Refs.~\cite{Sil73, Kom05, Kat10, Blu17B} for CR electrons and/or positrons, where $dE/dt$ represents ionization and synchrotron losses.

If the cross section is very large, the DM energy loss rate dominates over escape and (for electrons) Standard Model energy loss processes. Interestingly, in this case the solution simplifies again and is approximately
\begin{equation}\label{highDMSol}
N(E) = Q(E)\,T_{\rm loss}(E).
\end{equation}
If, for example, $T_{\rm loss}(E) \propto const.$, then $Q(E) \propto E^{-2.7}$ would be required to reproduce the observed spectrum, which would be a much softer injection spectrum than predicted by theory and inferred from gamma-ray observations of supernova remnants \cite{Bla13, HESSsurvey, Fermi13}.

Sufficiently large cross sections such that DM energy loss dominates over escape can be ruled out based on energy considerations alone. In addition, when $T_{\rm esc}$ and $T_{\rm loss}$ are comparable, DM-CR scattering can be probed by examining the distortions that DM scattering would induce in the observed CR spectrum. We discuss these two cases in the next two subsections.

\subsection{Conservative Limit From Total Energy Loss}\label{sec:energybudget}
Without assuming a detailed form for the source spectrum (as described above), we can obtain an extremely conservative limit by considering the source CR energy budget. In the Leaky Box Equation, if we fix $N(E)$ to be the measured data, any change in the energy loss rate due to DM must be compensated by a change in the source spectrum $Q(E)$. If supernovae are indeed the sites of CR acceleration, then in the absence of interactions with DM, $\sim 10\%$ of their energy must go into CRs to produce the observed CR spectrum \cite{Str10, Bla11, Der13, Bla13}. The larger we make the DM-proton cross section, the more energy must be injected into CRs to compensate for the larger energy loss. It is conceivable that the acceleration efficiency could be higher to compensate for energy loss due to scattering with DM, but the absolute most by which it could increase is clearly a factor of $\sim 10$. Because the uncertainty in this efficiency is the largest uncertainty in determining the total power injected into galactic CRs, we only require that the total power injected into CRs not increase by a factor of more than 10 compared to the best fit with no DM. We obtain an upper limit on the DM-proton cross section which becomes independent of mass for $m_{\chi} \ll  m_p$, as discussed in Sec.~\ref{sensitivity}, approaching $\sim$ 10$^{-25}$ cm$^2$ at around a keV. For the remainder of the mass range we consider, the limit curves upward due to a kinematic transition between the energy loss rate scaling as a constant and scaling as $m_{\chi}^{-2}$ (for $m_{\chi} \gg m_p$, outside our range of interest).

\subsection{Constraining $\sigma_{\chi p}$ with CR Proton Spectrum}

For a given DM mass and cross section, we compute the spectrum in the presence of DM using Eq.~(\ref{LBSol}). We assume that the cross section is independent of energy and velocity; in Sec.~\ref{discussion}, we discuss alternatives. We again fit over the seven parameters listed above: $\delta$, $\gamma_1$, $\gamma_2$, $\gamma_3$, the break energies $E_{B1}$ and $E_{B2}$, and the source spectrum normalization $Q$. To constrain the DM-proton cross section, we first compute the $\chi^2$ for the best fit with no DM. Then for a series of increasing cross section values, we compare the default $\chi^{2}$ to that when DM energy loss is introduced. Conservatively, we only exclude DM cases where, even allowing all CR parameters to vary in each step, the fit is worse than the one with no DM. We do assume a particular form for the no-DM spectrum, but the broken power law we use is sufficiently general and provides an excellent fit to the data.

The fit parameters in our model are not totally free (we discuss below how the results change if these restrictions are relaxed). First, we require that the source spectrum not be harder than $E^{-2}$, as predicted for diffusive shock acceleration in supernova remnants, thus \{$\gamma_1, \gamma_2, \gamma_3$\} \textgreater\:2.0. Second, as discussed above, the energy injected into CRs cannot be arbitrarily large. Rather than defining a hard cutoff, we penalize the fit for requiring a large normalization by defining a modified $\Delta \chi^2$:
\begin{equation}\label{deltachisqr}
\Delta \chi^2_{\rm mod} = \Delta \chi^2 + \frac{\textrm{Log}_{10}(\int dE \,Q(E)/\int dE \,Q_0(E))^2}{(\Delta Q)^2}\, ,
\end{equation}
where $\Delta \chi^2$ is the difference in $\chi^2$ value between the no-DM fits and the fits with DM for that particular DM mass and cross section. The second term constrains the total injected energy. We choose $\Delta Q = 0.2$, so that this term alone will contribute 25 to $\Delta \chi^2_{\rm mod}$ if the injected energy is 10 times larger than with no DM scattering. For our limit on DM-proton scattering, this additional source-normalization term is the dominant contribution to $\chi^2_{\rm mod}$ for $m_{\chi} \gtrsim 1$ MeV, but is unimportant for lower masses. We integrate from 1 GeV to 100 TeV, roughly the end of the data range we use.  The lower limit of integration is chosen to cover the bulk of the CR energy content.  Changing this limit of integration to 10 GeV weakens our results for $m_{\chi} \gtrsim$ 1 MeV by only a factor $\lesssim$ 2. The source spectrum break energies are left unconstrained, except for the requirement that they lie within the energy range we consider. And although there are measurements of $\delta$, to be conservative the only restriction we place is that $\delta > 0$, meaning that the escape time does not increase with energy, a physically well-motivated restriction.

For each DM mass, we find our limit by increasing the cross section in small steps, refitting all CR parameters on each step.  This $\chi^{2}$ value thus depends on the cross section alone, and monotonically increases with cross section (no preference for DM is found).  We increase the cross section until
\begin{equation}\label{chisqrlimit}
\Delta \chi^2_{\rm mod} = 25\,,
\end{equation}
which yields our 5$\sigma$ upper limit in the cross section. We choose to report 5$\sigma$ limits rather than more conventional 2$\sigma$ or 3$\sigma$ limits to be conservative, so that it is clear that we place robust limits on particle properties despite all astrophysical uncertainties.

\begin{figure}[t]
\centering
\includegraphics[width=\columnwidth]{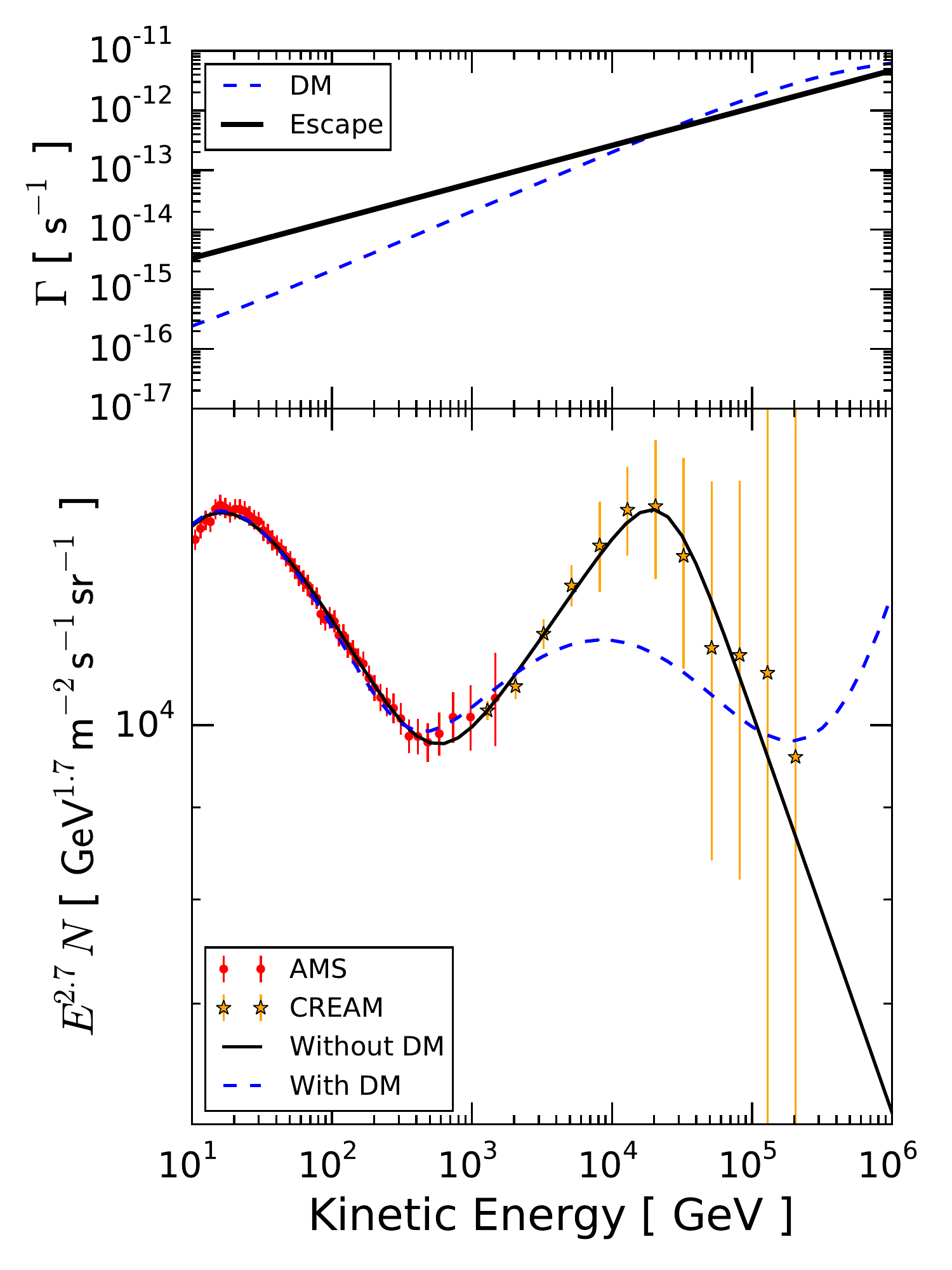}
\caption{Effects of DM on proton CRs, for $m_{\chi}$ = 1 keV and $\sigma$ = $2.0 \times 10^{-27}$ cm$^2$. {\bf Top:} Escape and DM energy loss rates. {\bf Bottom:} CR proton spectrum measured by AMS \cite{AMSproton} and CREAM \cite{CREAM}, with the best-fit spectrum with and without DM interactions. The cross section for the dashed curve has $\Delta \chi^2_{\rm mod} = 25$. Plots for other masses are in the Appendix.} 
\label{fig:badfit}
\vspace{-0.25 cm}
\end{figure}

Figure~\ref{fig:badfit} shows an example case that we consider to be ruled out to demonstrate how DM-proton scattering can affect the CR spectrum.  
For this example~($m_{\chi}$ = 1 keV and $\sigma = 2 \times 10^{-27}$ cm$^2$), we can see the relative importance of the DM energy loss by comparing the CR escape rate~($1/T_{\rm esc}$) and DM energy-loss rate~($1/T_{\rm loss}$). In this case, $1/T_{\rm loss}$ overtakes $1/T_{\rm esc}$ above $\sim$10\,TeV and becomes the dominant process; this extra energy loss produces an imprint in the final proton spectrum.  This effect can be seen by comparing the model spectrum with and without DM.  In this case, the extra energy loss above 10\,TeV suppresses the spectrum, modifying the overall broken power-law behavior, and thus results in a worse fit to the data. The restriction that \{$\gamma_1$, $\gamma_2$, $\gamma_3$\} \textgreater\:2.0 is invoked over much of the mass range we consider. In Sec.~\ref{discussion}, we discuss how our limit would change if we relaxed this restriction. 

\begin{figure}[t]
\centering
\includegraphics[width=\columnwidth]{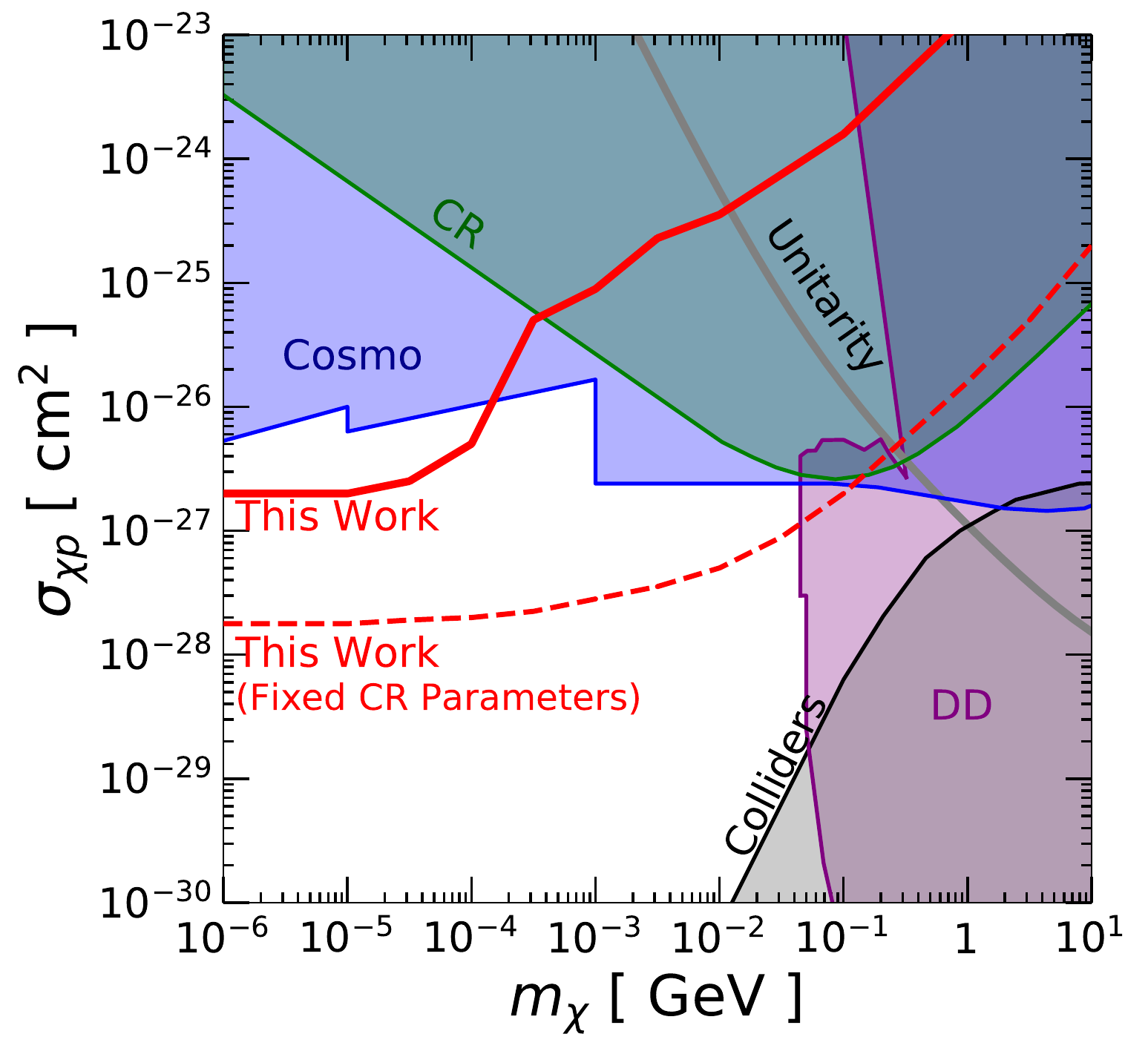}
\caption{Our limit on $\sigma_{\chi p}$ (solid red), along with the sensitivity that this method could achieve if the CR fit parameters were known precisely (dashed red). Our limit is compared to other limits from cosmology, CRs, colliders, and direct detection; see Fig.~\ref{fig:prevlims} for details. Gray is the unitarity bound for s-wave scattering at a CR energy of 10 GeV; see text for details.} 
\label{fig:proton}
\vspace{-0.25 cm}
\end{figure}

Figure~\ref{fig:proton} shows our limit on the DM-proton elastic cross section from 1 keV to 10 GeV. For $m_{\chi} \lesssim 1$ MeV, our limit is set by the spectrum shape. In this range, $T_{\rm Loss}$ is $\propto E$. Thus, if the cross section is large and energy loss dominates over escape, the solution for the spectrum at Earth will approach $E^{-(\gamma_i+1)}$. Since we require $\gamma_i > 2$, this cannot be harder than $E^{-3}$, and thus produces a bad fit to the data. On the other hand, for $m_{\chi} > 1$ MeV, $T_{\rm Loss}$ approaches a constant, so the data can be fit well with a soft injection spectrum. In this case, the limit is set by the energy budget consideration, which explains the weakening of the limit around 1 MeV. Overall, our method produces the tightest existing limit for masses below 100 keV.

As mentioned in Sec.~\ref{sensitivity}, for low $m_{\chi}$ it may take many collisions for a CR to lose an appreciable fraction of its energy. For $m_{\chi} = 1\,$ keV, at the cross section where we set our limit, a 10 GeV CR proton would scatter with DM $\sim10^4$ times as it propagates in the galaxy. For larger masses, going from 10 keV to 10 GeV in order-of-magnitude increments, the required numbers of collisions are \{$10^3$, $10^2$, $10^2$, 1 , 1, 1, 1\}. In our analysis, we treat energy loss due to collisions with DM as a continuous loss process. This is clearly appropriate in the low-mass case, in which many collisions, each producing small energy loss, are required. Our treatment may appear less valid in the high-mass case, where a single collision causes a proton to lose a substantial fraction of its energy. However, the differential equation we solve does not describe the propagation of a single particle, but the full spectrum of CRs, and hence the expected number of collisions is for the ensemble average, and fluctuations are small.

One way in which our method is conservative is that we allow the CR parameters to vary over large ranges. In principle, if some of the parameters can be determined by other methods, such as joint analyses of CR primary and secondary species, the limit can be significantly improved. To assess this, we repeat our analysis for each DM mass, but with all the CR parameters fixed to that of the no-DM case, leaving only the DM cross section free.  (Note that in this case $\Delta \chi^2_{\rm mod} \equiv \Delta \chi^2$ as $Q(E)$ is fixed to $Q_0(E)$.) The resulting limit is shown in Fig.~\ref{fig:proton}. With the CR parameters fixed, our limit is set by the shape of the CR spectrum at all $m_{\chi}$, and does not rely on the energy budget constraint. We again require that $\Delta \chi^2 = 25$, and consider this our best achievable limit using the selected CR data.  We find that this is one to two orders of magnitude stronger than our default result, meaning that better CR modeling may improve sensitivity significantly. 

Both our limit and our projected sensitivity approach constant values at low dark matter mass. This behavior can be understood by considering either Eq.~(\ref{eq:elossxsec}) or Eqs.~(\ref{DMloss}) and (\ref{deltaE}): in the low $m_{\chi}$ limit, the energy loss rate is independent of $m_{\chi}$, so the limit we set on the cross section is independent as well.

From Eq.~(\ref{deltaE}), it is apparent that the mass at which our limit flattens out is inversely proportional to the CR energy considered. In the future, if we had access to higher-energy CR data with small error bars, the sensitivity could be improved, with the flattening being shifted to lower mass.

\section{Constraining The DM-Electron Cross Section, $\sigma_{\chi e}$}
\label{electrons}

DM-electron scattering is less constrained than DM-proton scattering: only one cosmological study \cite{Ali15} has probed DM-electron scattering in the sub-GeV mass range, and existing direct detection experiments \cite{Ess12,SENSEI19,Agn18,SENSEI} are shielded by the atmosphere and Earth's crust, and are nominally sensitive only to masses above about 1 MeV. There should also be a ceiling to the collider search constraints, which, to our knowledge, has not been computed.

\subsection{Electron Data and Model Without DM}

Unlike the case for protons, Standard Model energy losses for electrons are non-negligible.  Therefore, even without the DM energy loss, we need to use Eq.~(\ref{LBSol}) to solve for the spectrum. From Ref.~\cite{Kom05}, we take
\begin{equation}\label{SyncIC}
\left(\frac{dE}{dt}\right)_{S+IC} \simeq 2\times10^{-16}\,  \left( \frac{E}{\textrm{GeV}}\right)^2\,\textrm{GeV}\,\textrm{s}^{-1}\,
\end{equation}
as the average energy loss rate due to synchrotron and inverse-Compton losses. As was done in the proton case for DM energy loss, we define a loss rate due to synchrotron and inverse-Compton effects as $|dE/dt|/E$. By comparing this loss rate to the same rate for DM and to the escape rate, we can see which effects are most significant at different energies.

We consider the CR electron spectrum from 10 GeV to 600 GeV, as measured by AMS \cite{AMSelect}. As for protons, we include statistical and systematic uncertainties, and treat the systematic uncertainties as uncorrelated, for simplicity and to be conservative. In principle, we could extend our analysis to higher energies by considering data from other experiments, such as DAMPE \cite{DAMPE17}, CALET \cite{CALET17, CALET18}, Fermi-LAT \cite{FLAT17} or HESS \cite{HESS09}. However, above about 200 GeV, there are significant discrepancies between the various data sets.

We model the CR electron source spectrum as a broken power law in rigidity, with spectral indices $\gamma_1$ and $\gamma_2$, break energy $E_B$, and normalization $Q$. We define $Q_0$ to be the value of $Q$ with no DM interactions, and will report the normalization required by a fit to the data in terms of the ratio $Q/Q_0$. We take the escape time to be a power law in rigidity, with exponent $\delta$, as in the proton case. The energy loss rate due to synchrotron and inverse-Compton losses is given above. Thus we have five fit parameters: $\gamma_1$, $\gamma_2$, $E_B$, $Q$, and $\delta$. Fitting this model to the AMS data, we find the best-fit values to be $\{ \delta, \gamma_1, \gamma_2, E_B, Q/Q_0 \} = \{0.6, 2.54, 2.29,120\, \rm{GeV}, 1.0\}$, all reasonable values. The resultant fit and the data are shown in the bottom panel of Fig.~\ref{fig:electronfit}. The $\chi^2$ per degree of freedom is 0.29, which is low, again likely due to our conservative choice to treat the systematic uncertainties as uncorrelated.  Overall, this shows that the CR electron data can be well-described by our simple CR model.

In our model with no DM scattering of CRs, the ratio between the differential source flux of electrons and protons is $\sim 0.01$ at 10 GeV, which matches the observed ratio at Earth \cite{bec05}, and is only weakly energy dependent. This ratio is found to be of order 0.01 through far-infrared and radio observations of starburst galaxies \cite{Lac10} and radio observations of M33 \cite{Sar17}. Observations of young supernova remnants \cite{Vol04, Ber06, Ber09a, Ber09b, Mor12} and simulations of electrons and protons in shocks suggest comparable or lower values \cite{Par05, Ell10, Lee13}, so we consider this ratio well constrained to be $\lesssim 0.01$. Therefore the only freedom we have to increase the normalization of the electron spectrum is the same, extremely conservative factor of 10 that we allow for the proton spectrum. So, as for protons, we allow the total energy in the electron spectrum above 1 GeV to increase by at most a factor of 10.

\subsection{Electron Spectrum with DM-Electron Scattering}

The DM energy loss rate for electrons, Eq.~(\ref{Eqa}), is
\begin{align}\label{elecloss}
\frac{dE}{dt} &=  c\,\frac{\rho_{\rm DM}}{m_{\chi}}\, \sigma\, \frac{m_{\chi}\left(2m_e E + E^2\right)}{(m_{\chi} + m_e)^2 + 2m_{\chi}E}\\
&\simeq c\, \frac{\rho_{\rm DM}}{m_{\chi}}\, \sigma\, \frac{E}{2} \,.
\end{align}
Unlike the case for protons, the DM-electron energy loss is always in the regime $2 m_{\chi}E \gg (m_{e}+ m_{\chi} )^2$.  Thus our method is more constraining at small DM mass.

\vspace{-0.5 cm}	

\subsection{Conservative Limit From Total Energy Loss}\label{sec:electron_energy}
We set a conservative limit on the DM-electron cross section by requiring that the total energy injected into CRs not increase by a factor of more than 10 to compensate for energy losses due to DM. The limit we obtain is $\sigma \lesssim 10^{-23} \left(\frac{m_{\chi}}{\textrm GeV}\right)\,{\rm cm^{2}}$. The mass scaling here is different than for protons because of the different mass dependence of the energy-loss rate.

\subsection{Constraining $\sigma_{\chi e}$ with CR Electron Spectrum}
To incorporate the effect of DM-electron scattering, we solve for the spectrum with DM energy loss using Eq.~(\ref{LBSol}), where now $dE/dt$ is the sum of the DM energy loss rate and the Standard Model energy loss rate.  At each mass, we fit over all the CR
 parameters for each cross section value, requiring as before that $\{ \gamma_1, \gamma_2 \} > 2$. We rule out cross sections with $\Delta \chi^2_{\rm mod} \geq 25$, where $\Delta \chi^2$ is defined as for the proton case, Eq.~(\ref{deltachisqr}).

\begin{figure}[t]
\centering
\includegraphics[width=\columnwidth]{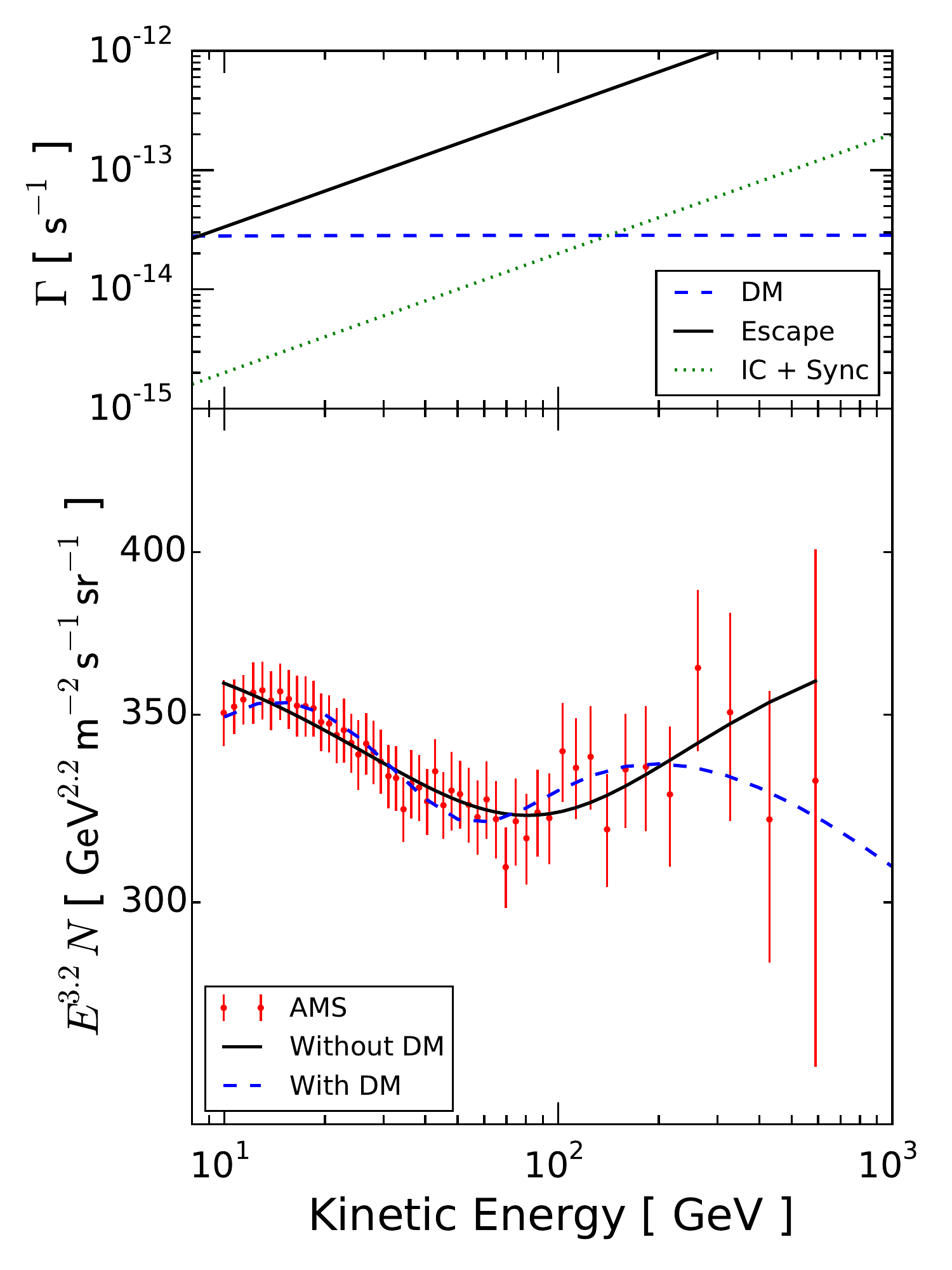}
\caption{Effects of DM on electron spectrum, for $m_{\chi}$ = 1 keV and $\sigma$ = $4.0 \times 10^{-30}$ cm$^2$. {\bf Top:} Escape, DM, and inverse-Compton plus synchrotron loss rates. {\bf Bottom:} CR electron spectrum measured by AMS \cite{AMSelect}, with a best fit spectrum with and without DM interactions. The cross section for the dashed curve has $\Delta \chi^2_{\rm mod} = 25$, which comes mostly from constraining the source energy budget~(see text for details). } 
\label{fig:electronfit}
\vspace{-0.25 cm}
\end{figure}

Figure~\ref{fig:electronfit}~(bottom panel) shows the AMS data, the best fit with no DM, and the best fit for an example case with $\Delta \chi^2_{\rm mod} = 25$ ($m_{\chi}$= 1\,keV, $\sigma= 4\times 10^{-30}$\,cm$^{2}$).  Unlike the case for protons, the restriction on the source spectral index, $\{ \gamma_1, \gamma_2 \} > 2$, is unimportant for fitting the electron spectrum. And because of the limited energy range considered, the curvature induced in the observed spectrum by scattering with DM is not significant enough to produce a bad fit to the data. Therefore, our constraint comes entirely from the energy budget, as our proton limit did for $m_{\chi} \gtrsim$ 1 MeV. As for protons, we include in $\Delta \chi^2$ the energy budget term $\frac{\textrm{Log}_{10}(\int Q/ \int Q_0)^2}{(\Delta Q)^2}$, where the integrals are carried out from 1 GeV to 600 GeV, the end of the electron data we use. As for protons, changing the lower limit of integration to 10 GeV weakens our limit by only a factor $\lesssim$ 2.

We note, from examining Eq.~(\ref{elecloss}), that the energy-loss rate is approximately proportional to $\sigma/m_{\chi}$ for $1\, \textrm{keV} < m_{\chi} < 1\, \textrm{GeV}$, meaning that as one increases $\sigma$ and $m_{\chi}$ together, the DM energy-loss rate (and thus the effect of DM on the spectrum) does not change.  Thus the example case shown in Fig.~\ref{fig:electronfit} in fact applies to the full DM mass range that we consider through a $\sigma/m_{\chi}$ scaling.  This also explains the simple linear behavior of our constraint with the DM mass. 


\begin{figure}[t]
\centering
\includegraphics[width=\columnwidth]{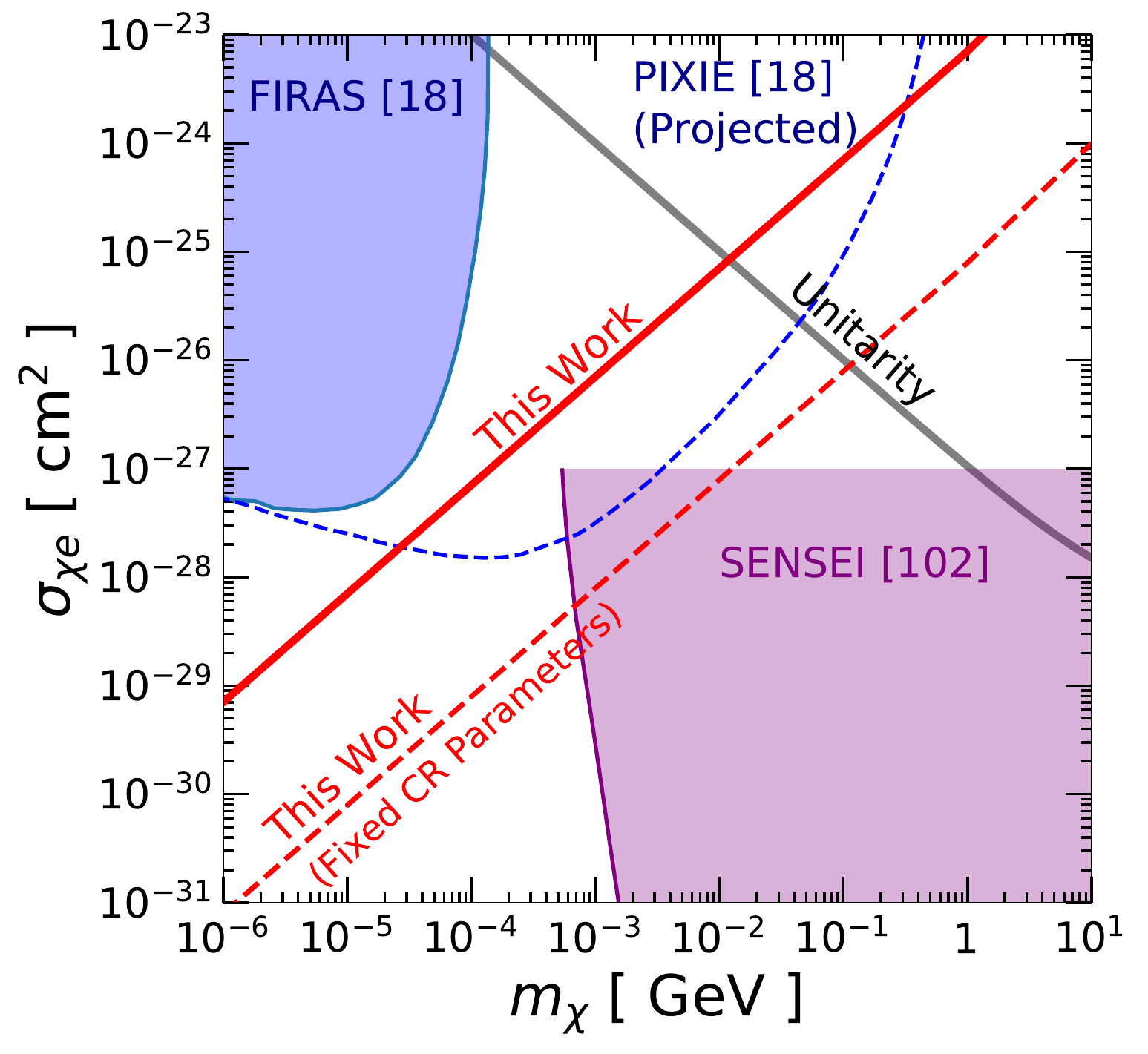}
\caption{Our limit on $\sigma_{\chi e}$ (solid red), along with the sensitivity that this method could achieve if the CR fit parameters were known precisely (dashed red). Also shown are existing and projected exclusion regions from cosmology, based on FIRAS and PIXIE \cite{Ali15}, and the existing exclusion region from SENSEI \cite{SENSEI19}. Gray is the unitarity bound for s-wave scattering at a CR energy of 10 GeV; see text for details.} 
\label{fig:electron}
\vspace{-0.25 cm}
\end{figure}

Figure~\ref{fig:electron} shows our limits on DM-electron scattering.  The constraint here is determined by the source energy budget.  From 1\,keV to 100\,keV, our constraint is better than cosmological constraints set by FIRAS~\cite{Ali15} by one to two orders of magnitude.  Above 100\,keV, our constraint covers parameter space that has not been probed. At the cross section where we set our limit, the average 10 GeV CR electron scatters $\sim 10\,$ times with DM as it propagates in the galaxy, independent of $m_{\chi}$.

The shape of our limit can be understood by considering the scaling of Eq.~(\ref{electronxsec}), or alternatively, Eqs.~(\ref{DMloss}) and (\ref{deltaE}). Over the entire mass range we cover, the energy loss rate is inversely proportional to $m_{\chi}$, so that the cross section must increase proportionally to $m_{\chi}$ in order to produce the same amount of energy loss. As in the case of our proton limit, the limit on the electron cross section approaches a constant in the low mass limit, but not until $m_{\chi} \ll 1$ eV. And as for protons, this break in the limit would move to even lower mass if we considered higher energy CRs.

In principle, direct detection experiments~\cite{Ess12,SENSEI,Agn18,SENSEI19} can probe the high-mass region.  In Fig.~\ref{fig:electron}, we show the exclusion region from SENSEI \cite{SENSEI19}; similar exclusion regions published by XENON10 \cite{Ess12} and SuperCDMS \cite{Agn18} have been superseded by the most recent SENSEI limits, so for simplicity we do not plot them. One challenge for direct detection experiments is that there is a ceiling to their exclusion regions due to the overburden. This ceiling is not trivial to calculate, so we simply show the exact exclusion region published by SENSEI. We do note that a ceiling for the XENON10 sensitivity region has been computed under the assumption of a dark photon mediator \cite{Emk17}, and is significantly below the top of the SENSEI region we show. Above these ceilings, more careful analyses are required to assess the sensitivity.  This highlights the importance and complementarity of cosmological and astrophysical probes to direct detection experiments.

Finally, we estimate the reach of our method with the AMS data in an optimistic scenario.  Similar to the exercise we perform with protons, we find the cross section that yields $\Delta \chi^2 = 25$, but with all the other CR parameters fixed at the no-DM case values.  In this case, the limit would be purely driven by distortion in the spectrum.  As shown in Fig.~\ref{fig:electron}, we find that with more careful modeling and analysis, there could be in principle a factor of $\sim$100 improvement to the sensitivity, without even considering additional CR data.

\section{Directions for Further Work}
\label{discussion}
In this work we set strong new constraints on \emph{relativistic} DM-proton and DM-electron scattering, as opposed to non-relativistic scattering probed in direct detection experiments. Our method is most sensitive at small DM mass and relatively large cross sections, which is a blind spot for typical DM searches and where the tightest constraints come from cosmology. For simplicity, and to be model-independent, we have also assumed energy-independent DM cross sections and isotropic scattering. Comparisons between our limits and cosmological constraints should thus be treated with caution, as cosmological limits typically become stronger relative to ours if the cross section decreases with velocity \cite{Bod18b}, while our limit becomes stronger if the cross section grows with energy (as considered in Ref.~\cite{Bod18}),making these two types of approach complementary.  While a detailed survey of DM models that are relevant in this regime is outside the scope of this work, we briefly comment these assumptions. Last but not least, we discuss ways to improve sensitivity.

One way for DM to have a large scattering cross section is to consider composite DM, with a geometric cross section comparable to that of a nucleus. In such a model, the s-wave cross section could conceivably be large, yielding an energy-independent cross section, as we have assumed. For example, the hadronic parts of nucleon-nucleon cross sections are fairly constant over several orders of magnitude in energy \cite{PDG}.

Another way to obtain a large cross section is scattering through a light mediator. A detailed examination of potential models is beyond our scope, but we outline some considerations to encourage future work. A useful analogy is $\nu_{\mu}+e^- \rightarrow \nu_{\mu}+e^-$ scattering mediated by the weak neutral current. When the center-of-momentum energy $E_{CM}$ is much greater than the mediator mass (in the Standard Model, that of the Z boson), the scattering cross section becomes energy-independent \cite{Gan98, For12}. For a 10-GeV electron scattering with DM of mass $m_{\chi}$ = 1 keV, a mediator with mass $M_{Z'} \ll \sqrt{2 m_{\chi} E_e} = \sqrt{2\, \rm{keV} \times 10\, \rm{GeV}} \simeq 5\, \rm{MeV}$ would produce a cross section that is both large (compared to our limits) and energy-independent. At lower $E_{CM}$, as appropriate for cosmological limits, for which the mediator would appear heavy, the cross section would scale with $E_{CM}^2$ (again, similar to neutrino-electron scattering).  If the cross section increases with energy, the constraints from CRs would effectively be much stronger than those from cosmology because, for the same couplings and mediator mass, the cross section would be larger at relativistic energies. However, scattering through a light mediator is heavily weighted toward forward scattering, and the energy loss rate given in Eq.~(\ref{DMloss}) would have to be modified to reflect this. Additionally, any model involving a light mediator would be subject to various constraints from astrophysics, cosmology, collider searches, and other terrestrial experiments \cite{Kna17}.

In general, DM scattering could also have cross sections that vary with energy in different ways.  If the cross section decreases or increases with energy, then the effect of DM energy loss would be shifted to the low or high energy part of the spectrum, depending on where the DM energy loss rate crosses with those of conventional processes.  In practice, we expect our results to be stronger if the cross section increases with energy, but weaker if the cross section decreases with energy, unless low energy data are included in the modeling. Additionally, the cross section for DM-CR scattering should be bounded above by a unitarity bound, such that $\sigma \lesssim \frac{4 \pi}{k^2}$, where $k$ is the momentum of each particle in the center-of-momentum frame \cite{Wei95}. In Figs. \ref{fig:proton} and \ref{fig:electron}, we show the corresponding unitarity bound for a CR energy of 10 GeV, the most relevant energy for our energy budget-based constraints. This bound gets lower for higher CR energy, but the section of our proton limit based on the shape of the spectrum, which relies largely on CRs in the 1--10 TeV range, still lies below the unitarity bound for 10 TeV CR energy. We also note that if there are strong features in the cross sections~(e.g., absorption peaks~\cite{Ng14, Blu14, Iok14}), they could leave imprints on the CR spectra as signatures of DM interactions.

In addition, we have neglected energy dependence introduced by the form factor of the proton. In our analysis, we do consider DM masses and CR energies where form factor suppression of the cross section may be large. However, in the mass range where our limit is competitive, and in the range of CR energies that matters for our limit, the effect of the form factor is negligible. Our limit on DM-proton scattering is most competitive for $m_{\chi} \leq 100$ keV. From Fig. \ref{fig:10-100kev}, it is apparent that the energy range that sets our limit at this mass is roughly 200 GeV \textless\, $E$ \textless\, 4 TeV. Using for example the electric form factor of the proton from Ref. \cite{Per06}, and an angular-averaged value of $Q^2 = \sqrt{2 E_{\chi} m_{\chi}}$, we find that the form factor suppression of the cross section is a factor of \{1.005, 1.4, 2.5\} for $E$ = \{200 GeV, 2 TeV, 4 TeV\}. Thus we find that the form factor suppression is only larger than 1.4 for two of the $\sim$ 12 data points that dominate the $\Delta \chi^2$, and thus negligibly affects our results. For 100 keV $\leq m_{\chi} \leq 1$ MeV, our limit is based on total energy loss, and is dominated by energies from roughly 1 to 100 GeV. For $m_{\chi} = 1$ MeV and $E = 100$ GeV, the angular-averaged suppression of the cross section is only a factor of 1.1. Finally, for $m_{\chi}$ \textgreater 1 MeV, form factor suppression starts to matter even for the energy budget constraint. But for these masses our limit is weaker than limits from cosmology and inelastic CR interactions by more than an order of magnitude, and also approaches the unitarity limit discussed above, so we neglect the form factor for simplicity.

Future observations will also determine with more certainty whether supernovae are in fact the dominant sources of both electron and proton CRs. If it were the case that CRs at energies we consider are accelerated with a spectrum much harder than $E^{-2}$, our limit on proton scattering with sub-MeV DM would get weaker and depend on the energy budget, as for higher mass. And if a class of sources with significantly higher luminosity than supernovae were determined to accelerate protons or electrons, some of our results would get weaker in proportion to the increase in total power. However, increasing the energy available to CRs would require a significant change to the current understanding of galactic CR acceleration. 

Another interesting direction to explore is inelastic scattering, e.g., pion production that leads to gamma rays, as considered in Ref. \cite{Cyb02}. In fact, our limit on DM-proton scattering is complementary: that limit is tighter than ours for masses above about 1 MeV, but loses sensitivity at low $m_{\chi}$ as the pion-production threshold increases, whereas because elastic scattering has no such threshold, our limit is tighter below 1 MeV. In principle, both elastic and inelastic scattering can be modeled together in the CR propagation framework, yielding a sensitivity that combines the strength of both processes.  But this will likely require a more model-dependent setup to specify how the inelastic interactions occur. We note that even below the pion production threshold, gamma rays could also be produced in DM-CR scattering via bremsstrahlung.  In principle, CR species other than protons and electrons can also be used to probe DM-CR interactions. DM-CR interactions could not only affect the energy of heavier nuclei, but could also contribute to their spallation interactions during propagation, and the threshold energy for spallation of nuclei is much lower than for pion production. Studies of CR elemental abundances could potentially be another sensitive probe of DM-CR scattering, especially at low mass where pion-production studies lose sensitivity due to threshold energy.

Improved modeling and tighter constraints on CR parameters would improve our sensitivity. As shown in Figs. \ref{fig:proton} and \ref{fig:electron}, precise knowledge of the CR source spectra and escape time would improve the sensitivity of our method by up to two orders of magnitude. A more detailed model of CR propagation, beyond the Leaky Box model, would take into account the CR source distribution and the DM spatial distribution in the galaxy. In particular, taking into account the enhanced DM density near CR sources (which are expected to be mostly in the inner part of galaxy) could further increase the sensitivity. The presence of DM-CR scattering may also decrease the level of CR anisotropies, observed to be $\sim 10^{-3}$ \cite{HAWC18}, which could be another avenue for constraining DM-CR interactions, but such a study would also require more detailed modeling than we do here.

Additional data would improve sensitivity as well. Given that the spectrum distortions produced by scattering with DM are broad in energy, modeling the CR spectrum over a wider energy range, and using future data with smaller error bars, would both make the shape of the observed spectrum more constraining. To extend the energy range, it would be important to take into account solar modulation in low energies, and to resolve the discrepancies between electron CR data at high energies. 
 
\section{Conclusions}
\label{conclusions}
We present novel tests of DM using reverse direct detection. Using CRs as the beam and DM as the target, we probe DM scattering with Standard Model particles. This approach, which examines the effects of DM-CR scattering on CR spectra, is sensitive to part of parameter space to which direct-detection DM searches are blind. We have shown that even with a simple model of CR propagation, a reasonable assumption about the source of CRs, and a conservative approach, existing data from AMS and CREAM can be used to set competitive limits on {\bf DM-proton} scattering, and to rule out large regions of previously unprobed parameter space for {\bf DM-electron} scattering. With more sophisticated modeling and analysis, precise measurements of the relevant propagation parameters, and inclusion of additional or newer data, the sensitivity could be significantly improved, potentially by a few orders of magnitude.

While we consider DM-proton and DM-electron scattering, other studies have also constrained DM-photon \cite{Boe14} and DM-neutrino scattering \cite{Man06, Arg17}. Together, these studies work toward the overarching goal of understanding DM's interactions with the Standard Model. By combining such techniques, future work will either discover DM or set overarching constraints on its interactions with all known particles.

\bigskip
\noindent {\bf Note added.}--- Shortly after our paper appeared on arXiv, Refs. \cite{Bri18} and \cite{Ema18} appeared, which consider the detectability of DM upscattered by CRs, which is complementary to our considerations of CR downscattering.

\vspace{1 cm}
\section*{Acknowledgments} 
We thank Kfir Blum, Kimberly Boddy, Antonio Boveia, Joseph Bramante, Ben Buckman, Timon Emken, Rouven Essig, Glennys Farrar, Brian Fields, Vera Gluscevic, Rafael Lang, Rebecca Leane, Kohta Murase, David Rainwater, Juri Smirnov, Todd Thompson, Aaron Vincent, and Edoardo Vitagliano for helpful comments and discussions. We thank Brian Fields in particular for the suggestion of studying cosmic ray anisotropies as a direction for future work, and Juri Smirnov and Edoardo Vitagliano for their help in understanding the unitarity bounds for dark matter-cosmic ray scattering. Finally, we thank the anonymous referee for helpful comments that improved the paper.
CVC and JFB are supported by NSF Grant No. PHY-1714479. KCYN is supported by the Croucher Fellowship and the Benoziyo Fellowship. 
			

\bibliography{reversedd}

\clearpage
\onecolumngrid
\section{Appendix: Additional Figures for DM-Proton Scattering}

\begin{figure}[h]
\centering
\includegraphics[width=.40\columnwidth]{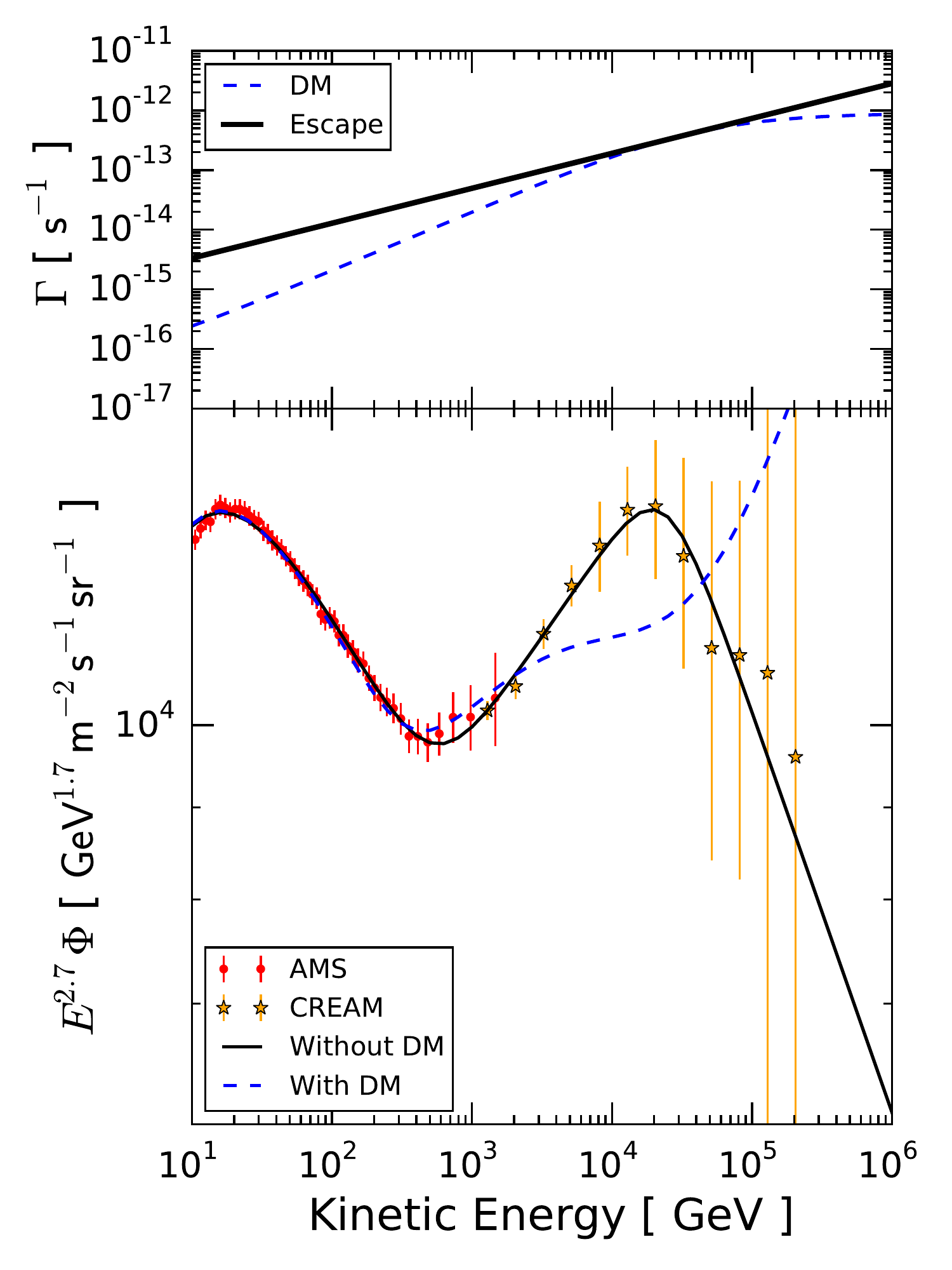}
\includegraphics[width=.40\columnwidth]{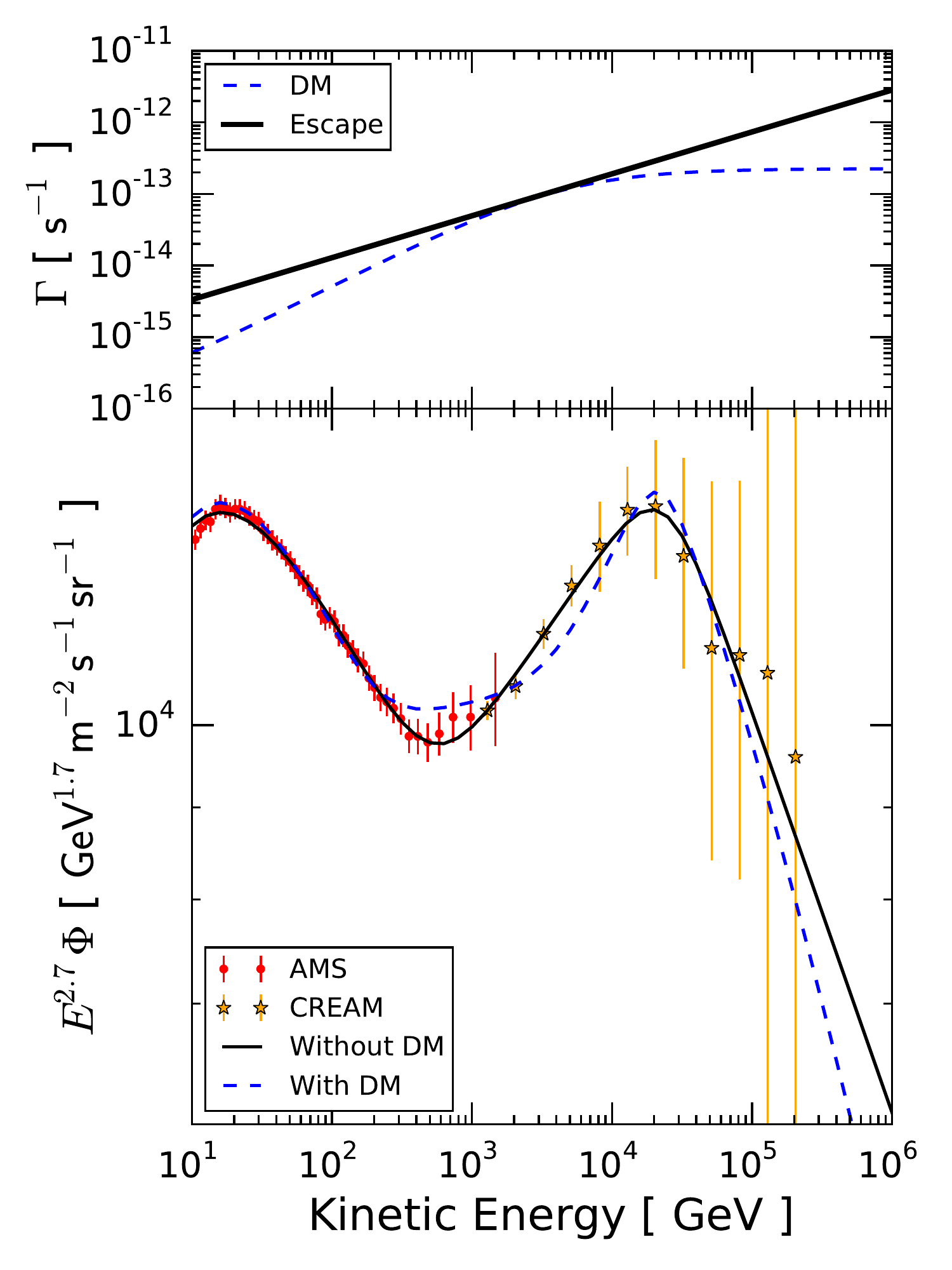}
\includegraphics[width=.4\columnwidth]{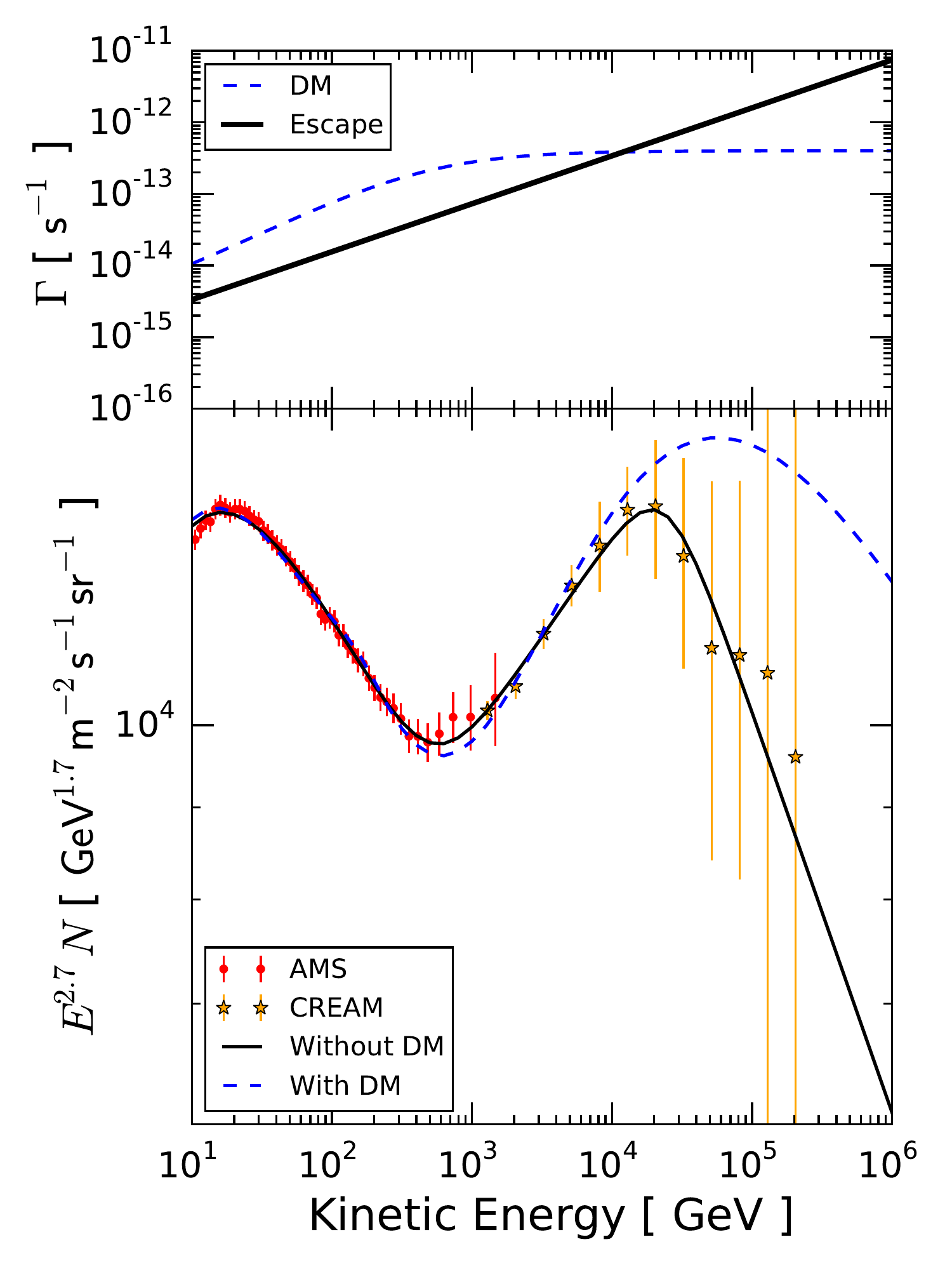}
\includegraphics[width=.4\columnwidth]{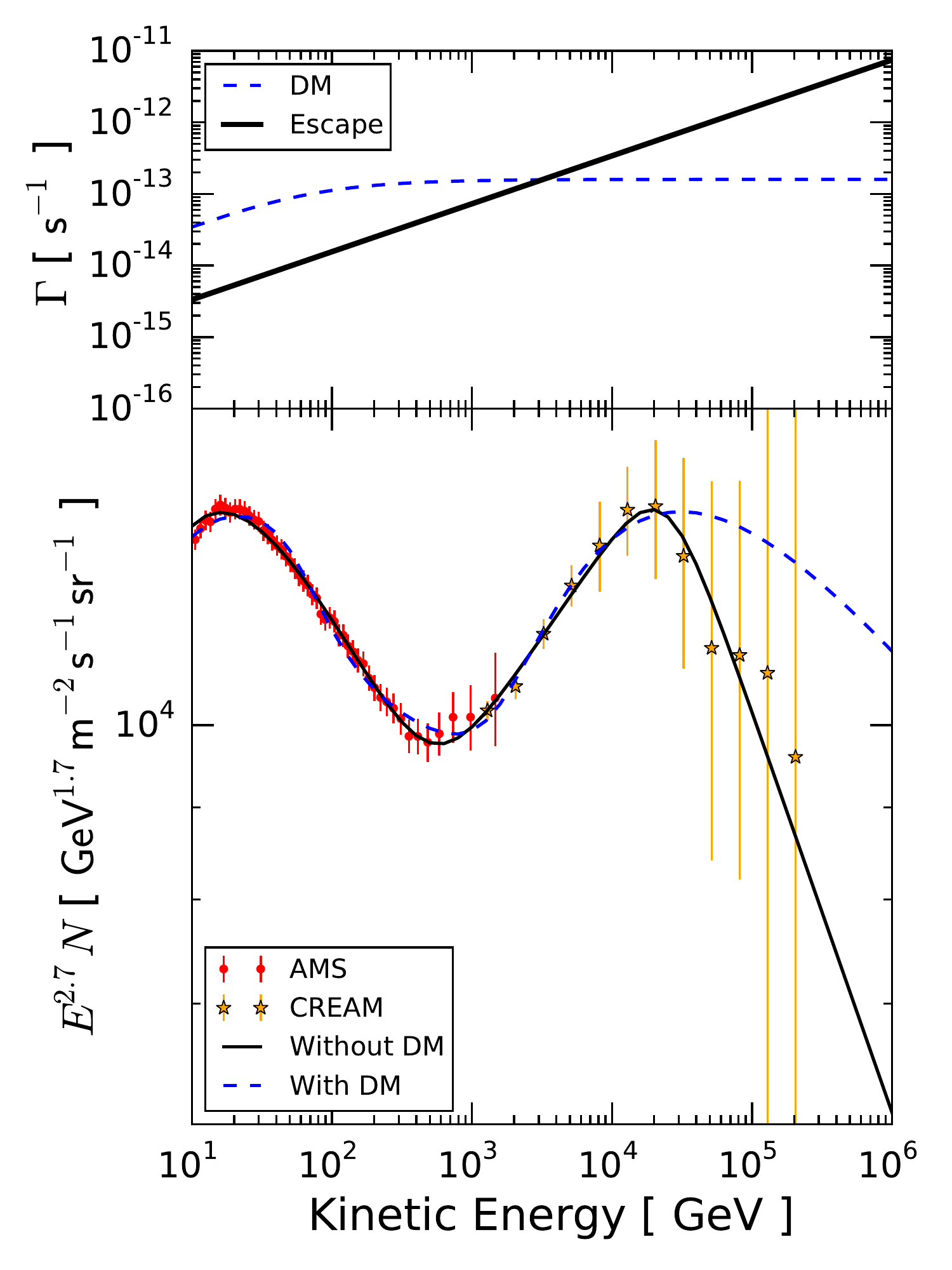}
\caption{{\bf Top left:} $m_{\chi}$ = 10 keV, $\sigma$ = $2.0 \times 10^{-27}$ cm$^2$. {\bf Top right:} $m_{\chi}$ = 100 keV, $\sigma$ = $5.0 \times 10^{-27}$ cm$^2$. {\bf Bottom left:} $m_{\chi}$ = 1 MeV, $\sigma$ = $8.9 \times 10^{-26}$ cm$^2$. {\bf Bottom right:} $m_{\chi}$ = 10 MeV, $\sigma$ = $3.5 \times 10^{-25}$ cm$^2$.} 
\vspace{-0.5 cm}
\label{fig:10-100kev}
\end{figure}

\begin{figure}[t!]
\centering
\includegraphics[width=.4\columnwidth]{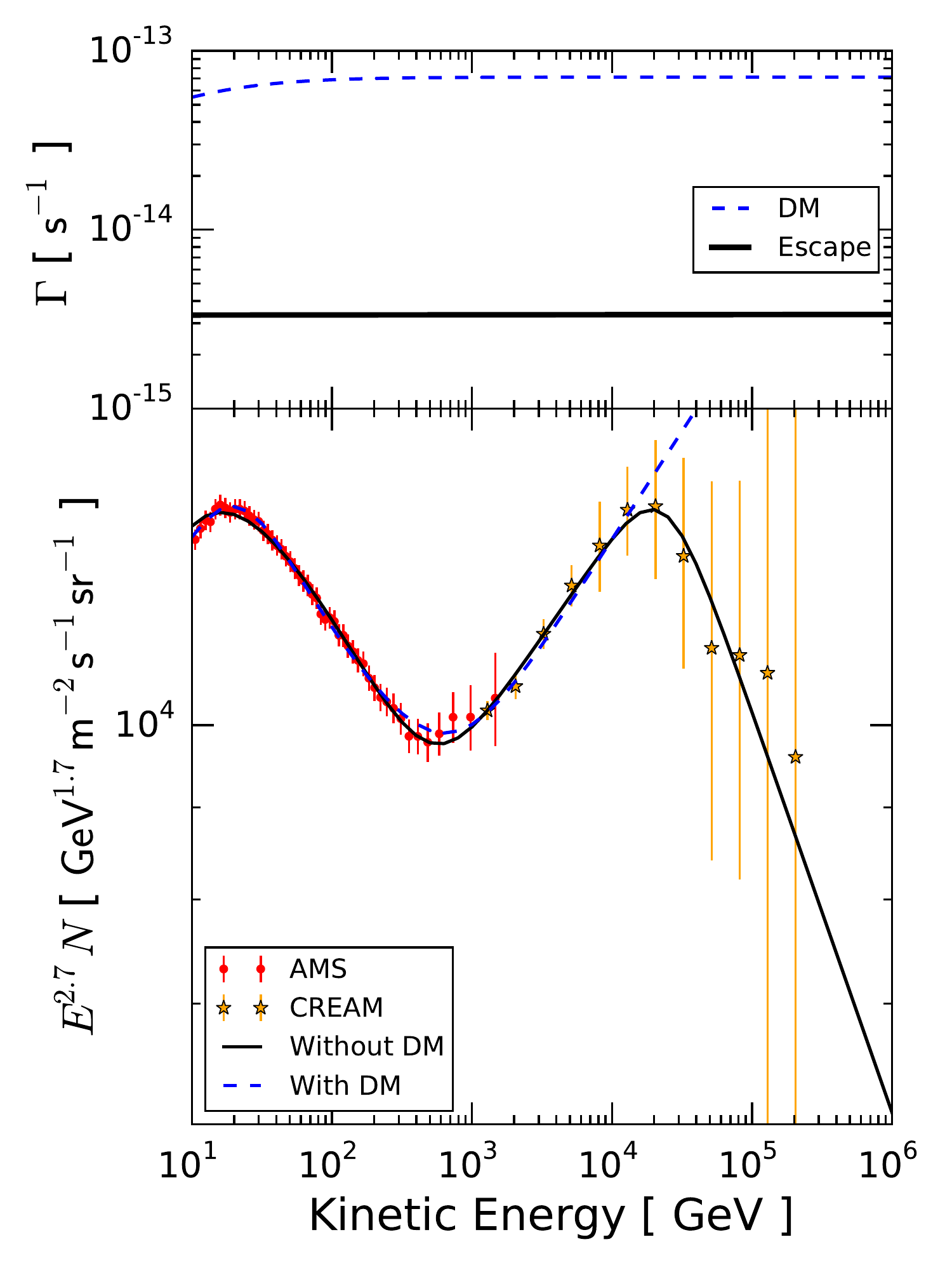}
\includegraphics[width=.4\columnwidth]{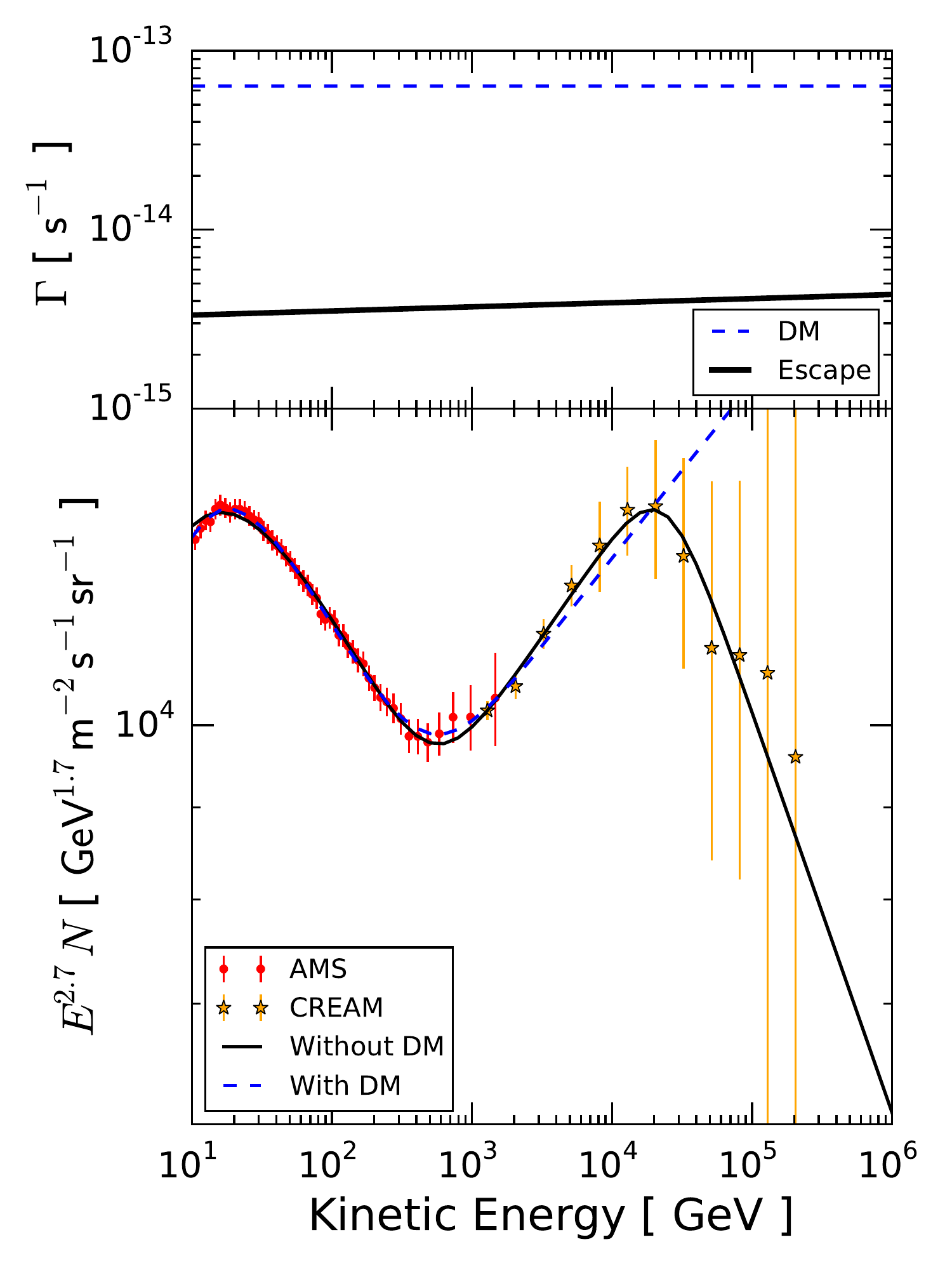}
\caption{{\bf Left:} $m_{\chi}$ = 100 MeV, $\sigma$ = $1.6 \times 10^{-24}$ cm$^2$. {\bf Right:} $m_{\chi}$ = 1 GeV, $\sigma$ = $1.4 \times 10^{-23}$ cm$^2$.} 
\label{fig:100mev-gev}
\end{figure}

Figures \ref{fig:10-100kev} and \ref{fig:100mev-gev} show the effects of DM-proton scattering on the CR proton spectrum for several DM masses~(analogs of Fig.~\ref{fig:badfit}). In some cases, the $\chi^{2}$ value is driven by the low-energy part of the proton spectrum, which is difficult to see in the wide-energy-range spectrum plots due to the high density of the AMS data.  For electrons, we do not show additional figures because different DM masses have the same effects that were shown in Fig.~\ref{fig:electronfit} due to the $\sigma/m_{\chi}$ scaling in the energy loss term.

\vspace{-0.5 cm}

\pagebreak

\twocolumngrid

\end{document}